# Habitat fragmentation affects climate adaptation in a forest herb

Local adaptation in *Primula elatior*

Frederik Van Daele[1], Olivier Honnay[1], Steven Janssens [2,3], Hanne De Kort[1]

1 Department of Biology, KU Leuven, Leuven, Belgium

2 Meise Botanic Garden, Nieuwelaan 38, BE-Meise, Belgium

3 Leuven Plant Institute, KU Leuven, Leuven, Belgium

Correspondence:

Frederik Van Daele

Frederik.vandaele@kuleuven.be

Abstract
1. Climate change and the resulting increased drought frequencies pose considerable threats to forest herb populations, particularly where additional environmental challenges jeopardize responses to selection. Specifically, habitat fragmentation may impede climate adaptation through its impact on the distribution of adaptive genetic variation, and cause evolutionary shifts in mating systems.
2. To assess how habitat fragmentation disrupts climate adaptation, we conducted a common garden experiment with *Primula elatior* offspring originating from 24 populations sampled along a latitudinal gradient with varying climate and landscape characteristics. We then quantified a range of vegetative, regulatory, and reproductive traits under distinct soil moisture regimes to evaluate imprints of local adaptation and phenotypic plasticity. Additionally, we conducted a more extensive field campaign in 60 populations along the same latitudinal gradient to evaluate the potential evolutionary breakdown of reciprocal herkogamy.
3. For large, connected populations, our results demonstrated an evolutionary shift from a drought avoidance strategy in southern populations to a drought tolerance strategy in northern populations. However, habitat fragmentation disrupted climate clines and the adaptive responses to drought stress in key traits related to growth, biomass allocation and water regulation. Additionally, our findings indicate the onset of evolutionary breakdown in reciprocal herkogamy and divergence in other key flower traits. The disruption of climate clines, drought responses, and adaptations in mating systems contributed to a substantially diminished flowering investment across the distribution range, with the most pronounced effects observed in southern fragmented populations.
4. *Synthesis:* We present novel empirical evidence of how habitat fragmentation disrupts climate adaptation and drought tolerance in a wide range of traits along the range of the forest herb *Primula elatior*. These findings emphasize the need to account for habitat fragmentation while designing effective conservation strategies in order to preserve and restore resilient meta-populations of forest herbs amidst ongoing global changes.

# 1. Introduction

Rapid climate change is increasingly disrupting long term imprints of natural selection on plant physiology and phenology (Becklin et al., 2016; Cleland et al., 2007). As climate change outpaces the evolutionary rate of local adaptation, population viability becomes increasingly compromised (Jump & Peñuelas, 2005). Habitat fragmentation, a result of rapid land-use changes, can exacerbate the impacts of climate change by constraining the ability of species to migrate and adapt to the changing environment (Honnay et al., 2002; Skov & Svenning, 2004). Furthermore, the independent and interactive effects of climate drivers and habitat fragmentation on local adaptations may have significant implications for plant viability and fecundity. Disentangling the effects of these global change drivers on local adaptations is essential in order to inform conservation and management strategies, such as assisted translocation and targeted gene flow.

Climate clines play a significant role in shaping local adaptation in plant species. As temperatures, precipitation, and other climatic variables differ across various geographic regions, plant populations have evolved to adapt to the local environmental conditions they reside in (Franks et al., 2014). These climatic adaptations can manifest as changes in genetics, physiology, phenology, or as phenotypic plasticity. For example, forest herbs evolved distinct strategies for reproduction (Graae et al., 2009), growth (De Frenne et al., 2011), phenology (Bucher et al., 2018; Cleland et al., 2007), resource allocation (Qi et al., 2019; Rosbakh et al., 2015), or investment in specialized leaf morphological or stomatal adaptations (C. Liu et al., 2018; W. Liu et al., 2020), driven by factors such as the annual growing degree days, cold tolerance, heat stress, and water availability (Skov

& Svenning, 2004). Phenotypic plasticity can also be relevant for the establishment and persistence of populations, especially in the face of increasing climate extremes and drought frequencies (De Kort et al., 2020). However, adaptive divergence between populations is often mediated by counteracting factors such as gene flow, genetic drift, biogeographical history, community dynamics, and interacting environmental pressures (Blanquart et al., 2013; Kawecki & Ebert, 2004).

Habitat fragmentation can lead to increased isolation, demographic stochasticity, and altered environmental and biotic interactions, all of which can assert distinct selection effects and impact the evolutionary dynamics of plant populations (Dubois & Cheptou, 2017). The adaptive capacity of species in response to climate change is thus influenced by complex eco-evolutionary feedback loops involving interactions between various global change drivers (Hanski et al., 2011). These feedback effects can be antagonistic, synergistic, or additive (Coors & DeMeester, 2009), resulting in unpredictable adaptive responses to climate change across a species' native range (Jump & Peñuelas, 2005). For example, the reduced genetic variation and increased inbreeding often observed in fragmented populations may lead to a reduced adaptive potential of plant populations (antagonistic effect), consequently jeopardizing adaptive stress responses (Leimu et al., 2010). In contrast, a lower gene flow can, in some cases, accelerate local adaptations by preventing the influx of mal-adapted genes or swamping of locally adapted genes, resulting in synergistic eco-evolutionary effects (Savolainen, Lascoux, and Merilä, 2013). Habitat fragmentation could also lead to increased variability in abiotic and biotic conditions, which could select for strategies that enable species to cope with heterogeneity (Cheptou et al., 2017). Furthermore, fragmented habitats are often characterized by distinct land-use histories, are subject to increased edge effects, which can lead to distinct microclimates, irradiation regimes, increased soil nutrient contents, and modified interspecific interactions (Baeten et al., 2015; Xiao et al., 2016). These altered microclimates and interspecific conditions in fragmented habitats could impose selection pressures on growth patterns (Blondeel et al., 2020; Zellweger et al., 2020) and could disrupt the phenology of spring blooming forest herbs (Kudo et al., 2008). Furthermore, non-climatic selection pressures in fragmented habitats could also disrupt climate adaptation due to pleiotropic effects (Armbruster, 2002; Etterson & Shaw, 2001; Strauss & Whittall, 2006). Since both habitat fragmentation and climate drivers can influence the evolution of plant traits, it is thus essential to disentangle how they interactively shape local adaptation and to determine to what extend habitat fragmentation disrupts the adaptive potential of plant populations.

Larger distances between remaining habitat patches and an unsuitable landscape matrix reduces the dispersal potential of seed propagules and decreased the potential for pollination, asserting evolutionary pressures on both dispersal and mating mechanisms (Cheptou et al., 2017; Opedal, 2019). A hostile landscape matrix increases dispersal risks and may result in selection pressures that prioritize reproduction within habitat fragments while decreasing seed dispersal investment (Bonte et al., 2012). On the one hand, limitation of the seed dispersal potential can lead to selection pressures which improve dispersal mechanisms in order to avoid inbreeding depression (Cote et al., 2017; Olivieri et al., 1995). Smaller plant population sizes, lower flower densities and increased spatial isolation disrupt plant–pollinator mutualisms in fragmented habitats (Aguilar et al., 2006; Eriksson & Ehrlen, 2001; Gómez-Martínez et al., 2020; Kwak et al., 1998). Altered or impoverished pollinator communities could lead to selection pressures for traits that improve pollinator attraction or co-evolutionary flower syndromes (Gumbert, 2000). For example, heritable local adaptations of "bull's-eye" patterns have been documented, in which the petal bases absorb UV light while the petal apices reflect it, effectively directing pollinators towards the reproductive structures (Koski & Ashman, 2013). Moreover, disruption of plant-pollinator networks in fragmented habitats can lead to increased selection pressures on herkogamy, the relative positioning of stigmas and anthers within flowers (Jacquemyn et al., 2012; Mitchell et al., 2008). Research indicates that noticeable evolutionary changes in herkogamy can rapidly emerge in response to severe disruptions in plant-pollinator networks (Opedal, 2019; Opedal et al., 2017). Nevertheless, the pace of this evolutionary process is expected to be more gradual in long-lived species that heavily rely on clonal reproduction (De Witte & Stöcklin, 2010). The evolutionary breakdown of herkogamy can lead to feedback loops, as reduced herkogamy can drive increasing levels of autonomous selfing (Herlihy & Eckert, 2007; W. Zhang et al., 2021). This may be beneficial when pollination is limited, but it can also increase inbreeding and reduce genetic diversity (Jacquemyn et al., 2012; Porcher & Lande, 2005).

Given the profound yet complex ways in which both habitat fragmentation and climate change can influence plant evolution, and considering the substantial global variation in habitat fragmentation levels, we designed our research study to disentangle their individual and interactive effects on the adaptive potential of plant populations. Specifically, we conducted a common garden experiment complemented by an extensive field survey to elucidate the impact of habitat fragmentation on the evolutionary trajectories of natural *Primula elatior* populations. Our study aimed to: (objective 1) quantify the phenotypic variation driven by climate adaptation across a north-south cline, including responses to periodic droughts; (objective 2) assess the extent to which habitat fragmentation disrupts climate adaptation, particularly under drought stress; and (objective 3) evaluate the effects of habitat fragmentation on evolutionary shifts in the mating system.

## 2. Methods

### 2.1. Study system and sampling

*Primula elatior* subsp. *elatior* (*Primulaceae* family) is a European forest herb with a native distribution range from southern France up to Northern Denmark (Taylor et al., 2008; Van Daele et al., 2021). *Primula elatior* is dispersal limited, self-incompatible and is regarded as a representative model species for the mainly slowly colonizing broadleaf forest herb layer (Van Daele et al., 2021; Verheyen et al., 2003). Rhizome-driven vegetative colonization is very limited (Baeten et al., 2015) and seed dispersal is mainly determined by short range mechanisms such as gravity (barochory) and wind (rolling anemochory). Occasional seed herbivory (endozoochory) or transportation by rivers (hydrochory) in riparian habitats occur as well (Endels et al., 2004; Vittoz & Engler, 2007). *P. elatior* generally requires seed dormancy to ensure that seeds only germinate under favourable conditions (Baskin & Baskin 2004; Taylor et al. 2008; Baskin 2014). This process typically involves a period of exposure to low temperatures, which is further modulated by changes in irradiation levels. Furthermore *P. elatior* has found to be especially sensitive to soil waterlogging and drought (Taylor et al., 2008; Whale, 1983). *P. elatior* has distylous flowers with reciprocal herkogamy and pollination is mainly performed by bumblebees (Dickinson, 1990).

To evaluate the interactive effects of climate and habitat fragmentation on the mating system and adaptive trait evolution, we collected seeds for a common garden experiment of *P. elatior* along a paired latitudinal gradient (Fig. 1; De Frenne et al. 2013) involving one highly fragmented and one well-connected forest patch (large population in contiguous habitat), ranging from southern France up to northern Denmark. The paired sampling design was chosen to disentangle landscape fragmentation from climate effects along the distribution range. Seeds were collected from 15 individuals in all 24 populations (12 pairs) originating from respectively southern (southern France), central (northern France up to the North of Belgium) and northern (Denmark) Europe (four pairs in each region). The number of individuals in each population were counted as a proxy for population size, and seeds were collected from widely spaced individuals. Herkogamy was determined in 60 populations along the paired latitudinal gradient with a total of 2116 evaluated individuals and three flowers per plant (n = 6896) in order to quantify the potential evolutionary breakdown of self-incompatibility.

### 2.2. Common garden

A total of ~30 seeds originating from at least 15 parents (F1) from the 24 populations were spread evenly in petri dishes filled with perlite and water for two treatments (i.e. with and without vernalisation), resulting in a total of 976 petri dishes. The effect of vernalisation was evaluated to monitor the effects of increasing minimum temperatures during winter times on germination success and germination lag time. In the vernalisation treatment, seeds were cooled at a maximum temperature of 6° C for 7 weeks. Germination was enabled under a controlled climate regime of 12 hours light at 21° C and 12 hours dark at 16° C (C. C. Baskin, 2014).

Seedlings were randomized in a controlled common garden greenhouse experiment with four offspring (F2) originating from each of 15 parents (F1) for each population (24) in two treatments (2880 individuals) around ~ the 28$^{th}$ of February in 2020. In a first treatment (control), the water availability neared field capacity while staying below allowable depletion to avoid drought stress (Appendix S1). Two drought periods were induced near the pre-determined wilting point (18.4%± SE 4.87% soil water content) in the second treatment to evaluate responses to drought stress, one during the autumn of 2020 (~25% soil water content) and one during the spring of 2021 (~37.5% soil water content).

## 2.3. Response variables

To evaluate evolutionary adaptation drivers related to plant establishment and phenology, we evaluated the germination lag time, the germination success, juvenile growth rate, and the flowering phenology. The germination lag time was evaluated every three days and the germination percentage was determined after 6 weeks (no new germination for a week). Following the germination experiment, seedlings were measured at the start of the common garden experiment (28$^{th}$ of February 2020). The length of seedlings was re-evaluated at the 21$^{st}$ of April in 2020 (after ~53 days)in order to determine the seedling growth rate (mm/day). Juvenile growth was reassessed the 13$^{th}$ of July in 2020 when plants reached adulthood (~83 days). Missing values of the seedling growth rate, juvenile growth rate, and overall juvenile growth (from seedling to adult) were assigned the population mean. As each of these variables had very similar growth rate patterns, they were reduced in dimensionality with principal component analysis (PCA) with varimax rotation. The first PC axis explained 78.74% of the variance with a sum of squared loadings of 2.36 and the three metrics had a high correlation with the first PC axis (r > 0.7). Flowering phenology or the flowering lag time was determined based on the cumulative degree days warmer than 4°C (CDD4) starting from 1 January due to its correlation to the canopy phenology (Baeten et al., 2015; Richardson et al., 2006).

To determine evolutionary adaptation drivers on plant persistence, we quantified the adult growth rate and resource allocation traits such as the root-shoot ratio and specific leaf area (SLA). The adult plant height was reassessed the 15$^{th}$ of June 2021 to evaluate the adult growth rate (~337 days) while resource allocation traits were determined at the end of the experiment (august 2021). Shoots were separated at the growing point and belowground biomass was rinsed with water to remove soil. Biomass was oven-dried for 48 hours at 70°C and the weight was determined with a precision of 10$^{-2}$ g. The specific leaf area (SLA; leaf area/leaf mass) was determined at the same time. First, the leaf area was calculated based on leaf scan images with the R package "Leaf Area v. 0.1.8" (Katabuchi & Masatoshi, 2015) and leaves were subsequently oven dried for 48 hours at 70°C and weighted with a precision of 10$^{-3}$ g.

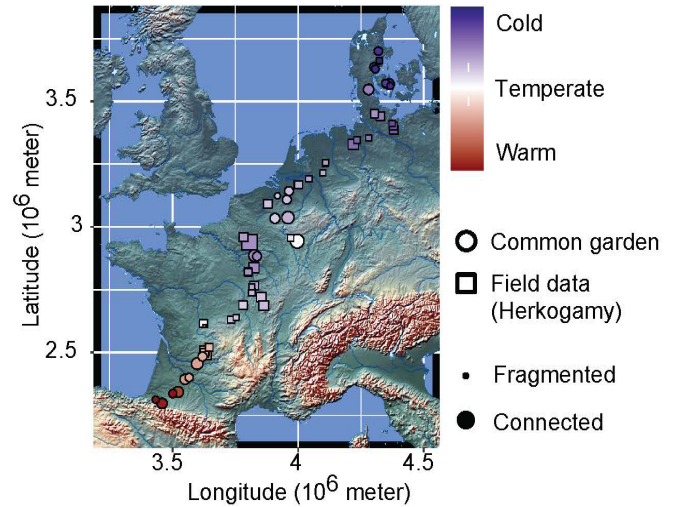

*Figure 1: Data collection locations of* Primula elatior *flowers and seeds. Herkogamy data was collected in all 60 locations. The seeds used for the common garden experiment originated from 24 of these locations.*

To evaluate micro-morphological leaf structure adaptations, we evaluated both abaxial and adaxial stomatal density, as well as the abaxial glandular trichome density. First, abaxial and adaxial leaf imprints were made with clear nail polish. For each imprint, three pictures were made with a Keyence VHX-5000 microscope magnified by 300x with a 688 x 516 μm surface area (0.36 mm²) for the adaxial side and 500x with a 1145 x 859 μm surface area (0.98 mm²) for the abaxial side. To classify and count the abaxial and adaxial stomata, and the abaxial glandular trichomes, custom convolutional neural network models were trained (Appendix S2), optimized and evaluated using "python v. 3.9.7" on the Flemish supercomputer network (Flanders Research Foundation, 2022) with the Tensorflow 2.8.0 (Abadi et al., 2015) and Keras 2.8.0 (Chollet & others, 2015) libraries, based on the guidelines outlined in the "from leaf to label workflow" (Meeus et al., 2020). To evaluate the neural network model quality, true positives (TP), false positives (FP) and false negatives (FN) were manually determined for a random stratified selection of 300 classified images for each model. Overall model performance was evaluated by the Jaccard index (i.e. threat score) of the binary confusion matrix: $\frac{TP}{TP + FP + FN}$. The threat scores of the CNN image classification models were 0.92 ± 0.07 SD, 0.73 ± 0.20 SD, and 0.81 ± 0.18 SD respectively for the abaxial stomata, adaxial stomata, and the abaxial glandular trichomes, indicating a high degree of similarity between the predicted and observed counts.

To evaluate the evolutionary adaptation drivers on mating mechanisms, which in turn determine the reproductive capacity, we determined the flower investment, bee view, herkogamy, and self-compatibility. Flower investment was determined based on the total count of produced flowers, the flower accessibility as determined by the mean flower opening duration (of three stalks), the total count of produced stalks, and the average stalk height. Missing values of these flower investment metrics were imputed with the population mean and reduced in dimensionality with PCA with varimax rotation and these variables were highly correlated with the first PC axis. To assess the local adaptation of visual pollinator signalling, we evaluated the three main bandwidths experienced by the main pollinators of *P. elatior*, namely bees from the genus *Bombus* (Schou, 1983). A macro-photography studio was constructed and natural light irradiation was standardised with a NISSIN MF-18 ring flash for the RGB bandwidth images, while a Convoy C8 365nm UVA with ZWB2 UV/VIS filter was used for UV bandwidth images. A custom algorithm was written for image segmentation, band separation, and false colour image construction using the R packages "magick v. 2.7.3" (Ooms, 2021), "imager v. 0.42.13" (Barthelme, 2022), "exifr v. 0.3.2" (Dunnington & Harvey, 2021), "hexView v. 0.3-4" (Murrell, 2019), "raster v. 3.4-13" (Hijmans, 2019), landscapemetrics v. 1.5.4" (Hesselbarth et al., 2019), SpaDES (Chubaty & McIntire, 2022), following the false colour bee view reconstruction recommendations of Lunau, Verhoeven and Ren (2018). The median flower colour value (0-255) of UV red, green and blue bands were reduced in dimensionality using a PCA with varimax rotation. The first PC axis explained 52.97% of the variance with a sum of squared loadings of 1.59. The first PC axis was 51.12% correlated to UV red variability, while the blue band was 86.1% correlated and the green band was 76.59% correlated. Stigma height (SH) and the anther height (AH) were separately determined in the field for 60 populations along the range (see study system and sampling) with a 0.01 mm precision digital caliper to evaluate herkogamy based on the guidelines of Brys and Jacquemyn (2015). The potential breakdown of biochemical self-incompatibility systems (Bawa et al., 2011; Dickinson, 1990) was evaluated with manual self-pollination to circumvent potential barriers caused by heteromorphic self-incompatibility. To this end, three flowers per plant (all plants in the control) were separately enclosed in string sealed tea bags prior to flower opening. During the reproductive stage, flowers were carefully self-pollinated with disposable cotton swab and re-enclosed in the sealed tea bags. Finally, when seed pods contained at least 1 developed seed they were scored as self-compatible (binary scoring of each seed pod).

## 2.4. Explanatory variables

To determine climate variability between populations, 19 bioclimatic variables (WorldClim version 2.1; Fick and Hijmans, 2017) were extracted in relation to the coordinates of the 60 habitat locations with R package "raster v.3.4-13" (Hijmans, 2019). A PCA with varimax rotation was used to reduce dimensionality and prevent multicollinearity with the R package "psych v.2.2.5" (Revelle, 2022). Based on the acceleration factor, which was determined with the R package "nFactors v.2.4.1" (Raiche & Magis, 2020), the first PC was retained as an explanatory variable. The first PC of the climate variables explained 64.21% of the variance and had a sum of squared loading of 12.20. The first PC displayed a strongly positive correlation with temperature and precipitation metrics and a strong negative correlation with temperature seasonality (Appendix S2).

Habitat fragmentation was evaluated based on a set of ecologically representable landscape metrics of broadleaf forest (Copernicus Land Monitoring Service, 2018) with a grid size resolution of 100m and the surveyed population size. The patch size (area metric), distance to edge (edge metric), and perimeter-area ratio (shape metric) were determined on a habitat patch level with the R package "raster v.3.4- 13" (Hijmans, 2019) and "landscapemetrics v.1.5.4" (Hesselbarth et al., 2019). To determine broadleaf forest class metrics, buffer radius distances of 2000m, beyond which little or no pollen and seed dispersal is expected were calculated from the patch centroid (Jacquemyn et al., 2002; Van Rossum, 2008). Within these landscapes, we calculated the mean patch area (area metric), aggregation index (aggregation metric; He, DeZonia and Mladenoff, 2001), cohesion (aggregation metric; Schumaker, 1996), division (aggregation metric; Jaeger, 2000), mean gyrate (area and edge metric; Keitt, Urban and Milne, 1997), number of patches (aggregation metrics), and the class percentage of the landscape (area and edge metric). These 11 metrics were reduced in dimensionality with PCA and a varimax rotation. Based on the acceleration factor, the first PC was retained as an explanatory variable. All metrics had a correlation coefficient (r) of at least abs(0.5) and the first PC explained 55% of the variance with a sum of squared loadings of 6.07 (Appendix S2). Lower values of this component represent higher levels of habitat fragmentation (decreased populations size, patch sizes, distance to edge, aggregation, cohesion, gyrate, and landscape proportion; and an increased perimeter-area ratio, division, and number of patches).

Vernalisation and drought treatments were indicated as binary factors to explain respectively the phenotypic plasticity of vernalisation on germination traits and phenotypic plasticity under drought stress of relevant common garden traits. Long-styled (L) and short-styled (S) flower morphs were incorporated as binary factors in the analysis of floral traits to better decipher the distinct evolutionary selection pressures on mating mechanisms associated with reciprocal herkogamy.

### 2.5. Modelling

To evaluate climate driven phenotypic evolution, including drought response adaptation (O1), and to quantify the disrupting evolutionary influences of habitat fragmentation on climate adaptation (O2), we modelled the three-way interaction of the first climate PC, the first PC of habitat fragmentation metrics, and the drought treatment or vernalisation factor (for germination traits) as drivers of adaptive trait divergence. Mixed models were constructed with in "lme4 v. 1.1-27.1" and model assumptions were evaluated and adjusted accordingly (Appendix S3) with the R package "performance v. 0.9.2" (Lüdecke et al., 2021) and "DHARMa v. 0.4.6" (Hartig, 2022).

The germination lag time was modelled with a negative binomial generalized linear model and log link function to control overdispersion of the deviating residual distribution with the R package "lme4 v. 1.1-27.1" (Bates et al., 2015). The germination success (rate) was modelled with a quasibinomial generalized linear model with a logit link to control for overdispersion, weighted by the total used seeds for each petri dish. The juvenile growth (PC1), the mean flowering phenology (CDD4; mean of three stalks), and adult growth rate were modelled with linear mixed-effects models with the parent ID (F1) as random factor. Models had a normal distribution of the residuals and random effect, homogeneity of the residual variance (homoscedasticity) was observed, and the model-predicted data was in line with the observed data. The root-shoot ratio (g) and SLA ($mm^2mg^{-1}$) was modelled with linear mixed-effects models with the parent ID (F1) as random factor. The root-shoot ratio was square root transformed and the SLA was log transformed to obtain a normal distribution of the residuals. The abaxial and adaxial stomatal density, and abaxial glandular trichome density was determined with linear mixed-effect models with the plant ID (F2) and parent ID (F1 origin) as nested random factors. For each model, the response was square root transformed to obtain a normal distribution of the residuals. The climate partition (phenotypic trait variation), partial effect of drought stress (phenotypic plasticity), and the contribution of their interaction (variability of phenotypic plasticity along the range) were evaluated in order to answer the first hypothesis (O1). Evolutionary processes (additive, synergistic, or antagonistic) driven by habitat fragmentation that disrupt climate adaptation (O2), were evaluated based on the contribution of the interaction effect of climate (PC1) and habitat fragmentation (PC1), and additionally by the contribution of the interaction effect of habitat fragmentation (PC1) and drought stress.

The extent to which habitat fragmentation drives evolutionary shifts in the mating system (O3) as determined by the flower investment was evaluated based on the partial habitat fragmentation contribution. The shifts in mating systems as determined by the bee view, herkogamy, and self-pollination was evaluated based on the habitat fragmentation partition, the morph type partition, and the interaction effect partition. To evaluate how habitat fragmentation drives evolutionary shifts in flowering investment, we first assessed the three-way interaction between climate PC1, habitat fragmentation PC1 and drought stress (PC1) with a linear mixed-effects model with the parent ID (F1) as random factor. The evolutionary shifts in bee view (pollinator attraction mechanism), herkogamy, and biochemical self-incompatibility was evaluated with the three-way interaction effect of the climate PC1, habitat fragmentation PC1, and the flower morph type. The three-way interaction effect on the mean bee view (mean of three flowers) was evaluated with a linear mixed-effects model with the parent ID (F1) as random factor. The mean herkogamy per plant was modelled with a generalized linear model and a gaussian error distribution with log link. Biochemical self-incompatibility was evaluated with a weighted binomial generalized linear mixed-effects model with logit link, the plant ID (F2) as random factor, and evaluated flower count as weights. Convergence failure was here addressed with the allFit optimizer algorithm in "lme4 v. 1.1-27.1" (Bates et al., 2015).

To partition the relative effects of climate and habitat fragmentation on trait variation in response to drought, we separately modelled each explanatory variable and their two-way interaction effects. The response to drought stress was replaced by vernalisation for the standardised variance partitioning of germination, while it was replaced by the morph type for the bee view, herkogamy, and biochemical self-incompatibility. Marginal effects of individual model terms were graphically represented using "sjPlot v. 2.8.9" (Lüdecke, 2021) and "ggplot2 v. 3.3.6" (Wickham, 2016) with habitat fragmentation as moderator variable using all observed values (Aiken & West, 1991; Dawson & Richter, 2006). Contributed proportions of the explained variance (Lüdecke et al., 2021; D. Zhang, 2021) of separate variables were based on their proportion relative to the three-way interaction effect. The contribution of two-way interactions were determined as contribution proportions relative to the three-way interaction effect minus the contribution proportion of both separate variables. Finally, the contribution of single effects and two-way interaction effects was standardised to eliminate potential deviations of the total contribution due to remnant confounding effects such as multicollinearity and suppression (Lai et al., 2022). The standardised variance partitions and statistical significance of separate terms and interaction terms were reported in venn diagrams using "VennDiagram v. 1.6.20" (Chen, 2018). Independent or interactive fixed effects on phenotypic trait variation were considered in detail when the effect was significant (p>0.05) and had a considerable contribution to the overall explained fixed effect variance (>5%), as determined by the variance partitioning.

We utilized redundancy analysis to examine the unique and shared influences of habitat fragmentation and climate on multivariate phenotypic trait variation and phenotypic plasticity (drought responses) using the vegan package version 2.5-7(Oksanen et al., 2013). First, values were averaged on the parent ID level (F1) and treatments were spread into a wide format to create separate phenotypic plasticity variables using "tidyr v. 1.1.3" (Wickham, 2021) while missing values were imputed with the population mean using "dplyr v. 1.0.7" (Wickham et al., 2021). Phenotypic plasticity variables were determined by the trait parent mean (F1) under control environmental conditions minus the trait parent mean under drought stress conditions. The redundancy analysis was modelled with the traits scaled to unit variance as response and the climate PC1 and habitat fragmentation PC1 as explanatory variables. Furthermore, phenotypic trait variation and phenotypic plasticity under drought stress variables were evaluated with cluster analysis. A dissimilarity matrix was calculated using "cluster v.1.2.1" (Maechler et al., 2017) with gower distances (Gower, 1971) and distinct clusters were delineated with hierarchical cluster analysis using the ward.D criterion (Murtagh & Legendre, 2014) and a three group threshold (height = 1). Increasing the groups (or reducing the height) further led to nested groups and reduced the overall ecological interpretability. Finally, the independent contribution of climate and habitat fragmentation was evaluated with variance partitioning using two partial redundancy analyses with respectively climate PC1 and habitat fragmentation PC1 as conditional variables.

## 3. Results

There was considerable climate driven trait variation along the range (O1). Vernalised northern seeds responded faster to temperature (cumulative degree days warmer than 4°C) compared to southern populations (Fig. 2; Appendix S4). When seeds were not vernalised, northern populations germinated significantly slower. In general, the germination rate was significantly higher in southern populations compared to northern populations. As expected, juvenile growth (PC1) was significantly faster in southern populations compared to northern populations. Even though drought stress only explained a minor fraction of the explained variation, it significantly reduced juvenile growth rate in northern populations while it decreased in southern connected populations (large southern populations in contiguous habitat). Flower phenology was significantly more sensitive to temperature cues (CCD4) in southern populations compared to northern ones, which resulted in faster flower development (Fig. 2). Under drought stress, northern populations flowered significantly faster while for southern populations flower development slowed down.

During the adult life phase, most of the explained variation in growth rate was driven by suppressed growth under drought stress and this was most pronounced in southern populations. The root-shoot ratio steeply increased in southern connected populations, indicating an increased investment in below-ground biomass (Fig. 3). However, under drought stress, we observed a decrease in the root-shoot ratio in individuals originating from southern populations and an increase in those from northern populations, specifically in landscapes unaffected by fragmentation (connected). The specific leaf area was significantly lower in southern populations.

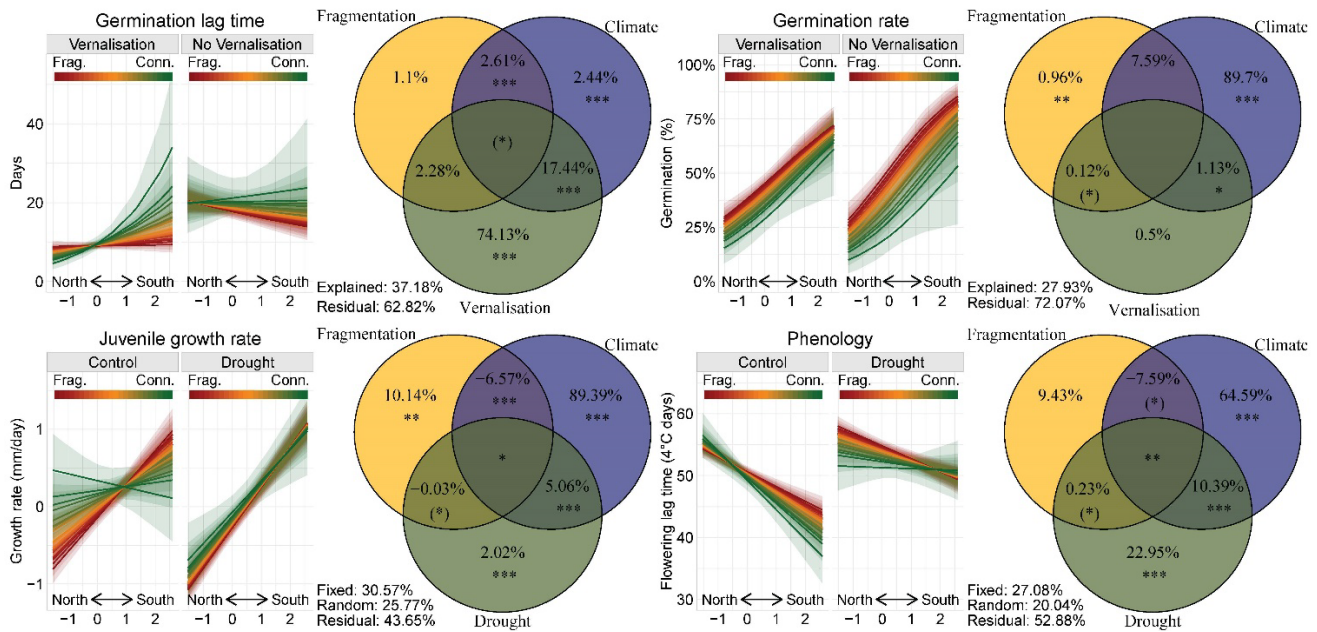

*Figure 2: The three-way interaction effect of climate, habitat fragmentation and drought stress frequency (or vernalisation during germination) on early life phases and phenological processes of* Primula elatior*. The germination lag time was expressed as cumulative degree days warmer than 4°C. The germination rate, here expressed as the germination percentage, is based on the germinated count relative to the total seeds (model weights) per petri dish as determined by the quasibinomial model. Juvenile growth rate displays the PC eigenvalues of the dimensionally reduced and imputed seedling growth rate, juvenile growth rate, and overall juvenile growth (from seedling to adult). The flowering lag time as an indicator of phenology displays the cumulative degree days (>4°C).*

However, the discrepancy between northern and southern populations decreased under drought stress.

The abaxial stomatal density was not significantly influenced by climate along the range (Fig. 4). Under droughts stress, however, southern populations increased the abaxial stomatal density while it decreased in northern populations. Adaxial stomatal density, on the other hand, did not display phenotypic plasticity and linearly decreased towards the south. The abaxial trichome density was slightly higher in northern populations and this increased further under drought stress.

Habitat fragmentation interfered with climatic clines (O2) for several traits (i.e. significant two-way interactions between habitat fragmentation and climate), and for most drought responses (i.e. two-way interactions between fragmentation and drought stress; and the three-way interactions between climate, habitat fragmentation and drought stress). Specifically, we found that the juvenile plant growth rate was relatively stable across the range for connected populations but was lower for fragmented populations in the north and higher in the south of *Primula elatior*'s distribution range. However, it has to be noted that even though the interaction between climate and habitat fragmentation was highly significant, the partial contribution of the interaction effect to the explained variance of the model was lower than the contribution of climate and habitat fragmentation separately (hence the negative contribution).

The flowering lag time was higher in southern fragmented populations, while northern populations flowered significantly slower under drought stress. Habitat fragmentation significantly limited adult growth in southern populations and increased adult growth in northern populations, thus nullifying the climate cline. However, the contribution to the explained variance was limited. The root-shoot ratio increased from north to south as an expected evolutionary response to warmer/drier conditions, but this adaptive response was absent across fragmented populations. Moreover, the plasticity response of the root:shoot ratio were inverted in plants from fragmented populations compared to plants from connected populations. Phenotypes from connected populations evolved higher plasticity towards drought stress with respect to abaxial stomatal densities, but this adaptive drought response was absent across fragmented populations. Additionally, fragmented populations displayed a reduced abaxial glandular trichome density in southern fragmented populations.

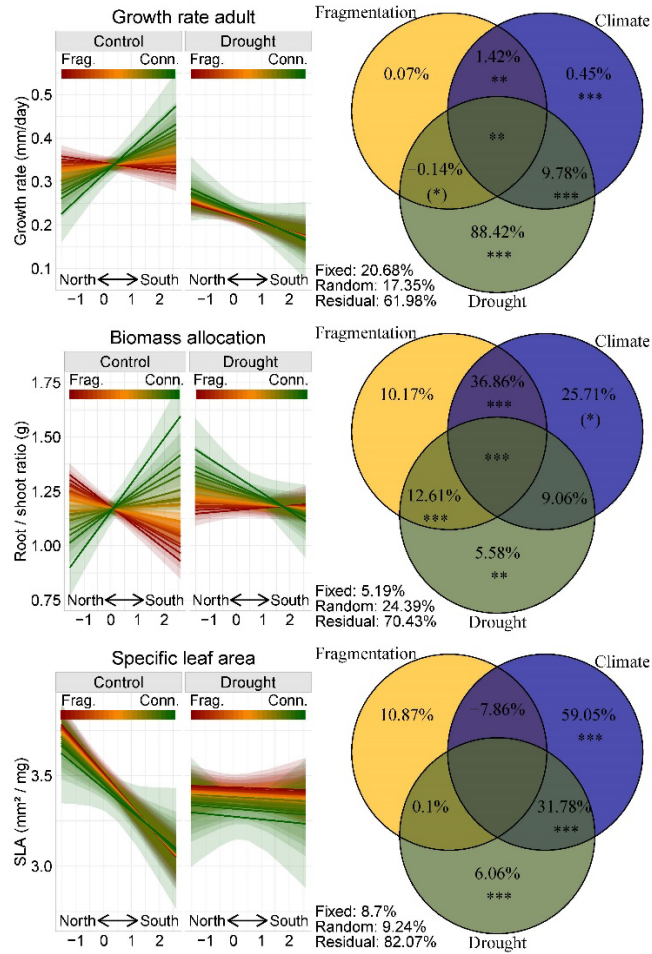

*Figure 3: The three-way interaction effect of climate, habitat fragmentation and drought stress frequency on adult growth and biomass allocation strategies of* Primula elatior. *Three-way interactions were predicted with climate PC1 as continuous variable, habitat fragmentation as moderator variable using all observed values (ranging from red to green), and drought treatment as binary categorical variable. The venn diagrams display the standardised contribution of fixed explained variance, as determined in the variance partitioning. The adult growth rate displays the growth in mm per day from 83 days up to 337 days of plant development. The biomass allocation displays the ratio between the dried root biomass divided by the dried shoot biomass in grams. The specific leaf area displays the leaf area divided by the leaf mass in mm²/mg.*

Flower investment was strongly reduced in fragmented populations along the range, being most pronounced in southern populations (Fig. 5). Pollinator signalling (bee view), herkogamy, and selfing success was not primarily driven by habitat fragmentation (O3), but rather showed substantial north to south variation (Fig. 5). The first PC values of the bee view increased in the south of the distribution and this adaptation was significantly more pronounced in fragmented habitats, which indicates an overall higher median flower reflectance of colour bands most visible to the clade *Anthophila* (blue, green, UV red). However, the large differentiation of flower reflectance between fragmentation levels was not apparent in northern populations. In addition, S morphs had a significantly higher overall reflectance than L morphs. Herkogamy, as an indicator of heteromorphic self-incompatibility, significantly increased in the south and this effect was strongest in S morphs from connected populations. However, S morphs originating from fragmented populations had a significantly lower herkogamy over the whole range and the divergence was especially pronounced among southern populations.

Biochemical self-compatibility was significantly higher in northern S-morphs close to the distribution edge, and this effect was most pronounced in fragmented populations. Succesful self-pollination and increased inbreeding coefficients in resulting offspring as compared to outcrossed offspring pointed to a successful breakdown of biochemical self-compatibility (Appendix S5). However, even though habitat fragmentation explained a high amount of variation (34.21%), it was not possible to discern a statistical trend due to a high variability between populations and also because connected populations at the northern edge of the range displayed an overall higher selfing rate.

Climate (F = 86.77, p = 0.001) and habitat fragmentation (F = 6.83, p = 0.001) explained 17.73% of the total phenotypic trait variation of which 16.27% was explained by climate (climate PC1 | habitat fragmentation PC1), 1.13% by habitat fragmentation (habitat fragmentation PC1 | climate PC1), and 0.33% of the variation could not be separated between both drivers. Climate was strongly aligned with the dissimilarity-based trait distances with plasticity of the flowering lag time and juvenile growth rate as strongest outliers (Fig. 6). Even though there was considerable dissimilarity between the individuals with the same parental origin (F1) along fragmentation levels, trait dissimilarity was considerably less with lowering investment (V) as strongest outlier.

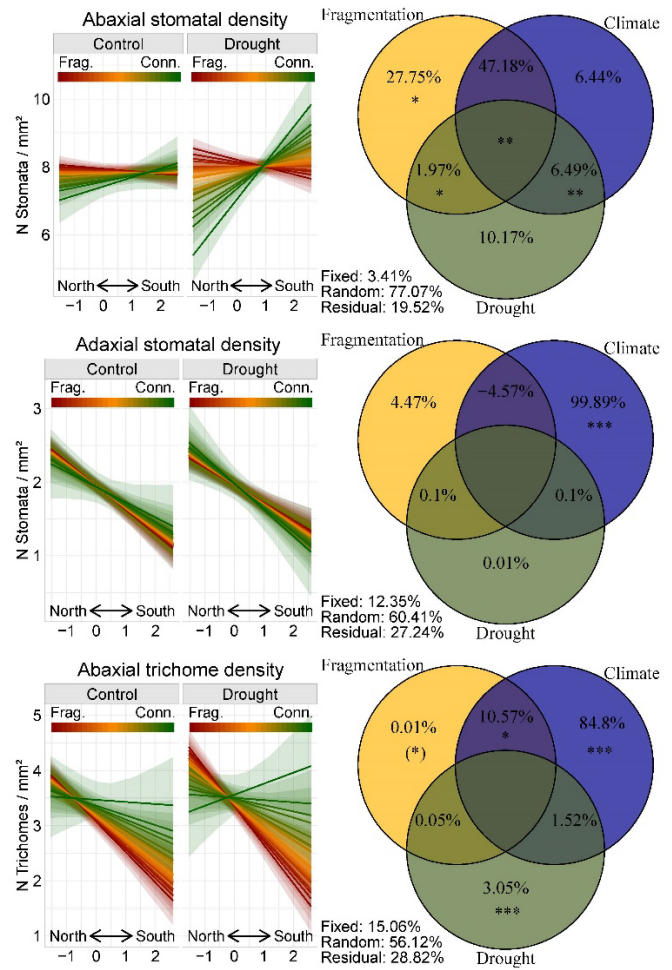

*Figure 4: Three-way interactions of climate, habitat fragmentation and drought stress on the abaxial and adaxial stomatal density, and the abaxial trichome densities of* Primula elatior. *Three-way interactions were predicted with climate PC1 as continuous variable, habitat fragmentation as moderator variable using all observed values (ranging from red to green), drought treatment as binary categorical variable. The venn diagrams display the standardised contribution of fixed explained variance, as determined in the variance partitioning.*

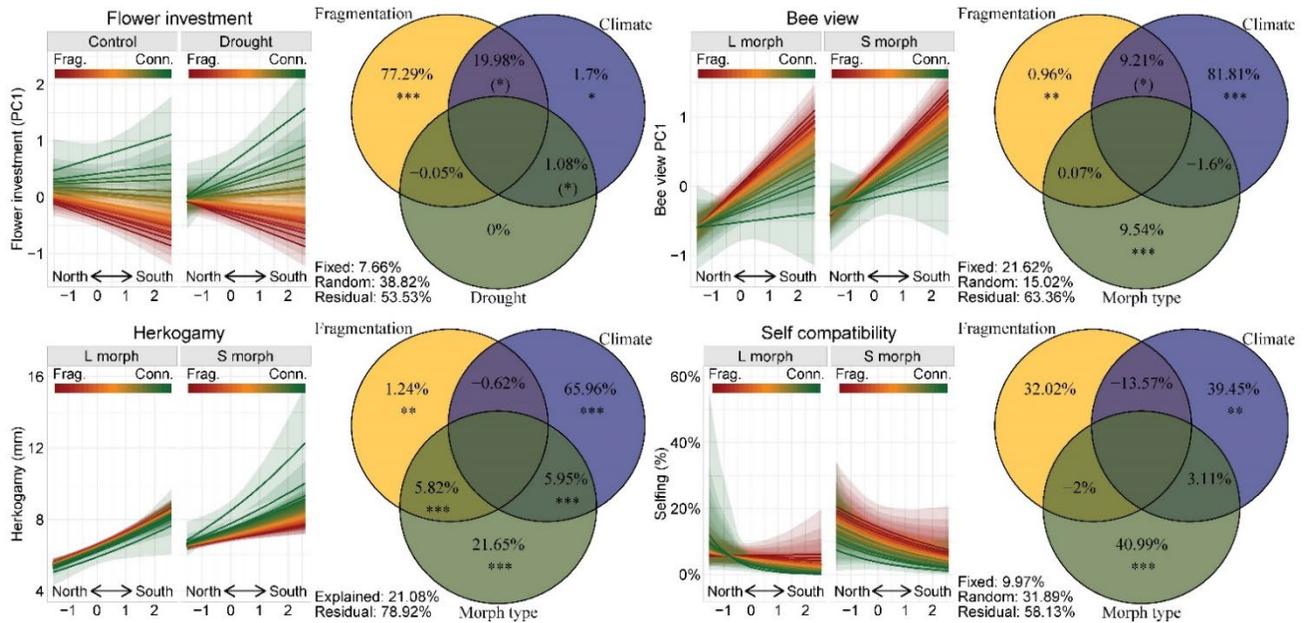

*Figure 5: Three-way interactions of climate, habitat fragmentation and drought stress on flower investment, bee view, herkogamy, and bio-chemical self-incompatibility of* Primula elatior*. The three-way interaction was predicted with climate PC1 as continuous variable, habitat fragmentation as moderator variable using all observed values (ranging from red to green), and drought treatment as binary categorical variable. Flower investment displays the first PC axis of the imputed total count of produced flowers, the mean flower duration (of three stalks), total count of produced stalks, and the average stalk height. The bee view displays the first PC axis of the median flower colour value (0-255) of UV red, green and blue bands, with higher values representing increased reflectance. Herkogamy was based on a separate large field measurement dataset (60 populations) and displays the absolute distance between anther and stigma. Self-incompatibility displays the relative percentage of selfed flowers that produced seeds. The venn diagrams display the standardised contribution of fixed explained variance, as determined in the variance partitioning.*

## 4. Discussion
### 4.1. Climate adaptation

Our results demonstrate considerable climate adaptation across the range of *P. elatior* in distinct life phases and traits (Fig. 6), which in turn determines plant performance and population viability (Lortie & Hierro, 2022). The potential for early plant performance is primarily driven by seed dormancy, lag time (initiation), germination success, and early growth strategies. Here, we found that seeds from northern populations were more sensitive to vernalisation (less cumulative degree days, warmer than 4°C, before germination) compared to southern populations (Fig. 2), indicating evolution towards increased seed dormancy in colder climates in order to avoid frost damage while maximizing the growing season (Bewley et al., 2012). The observed differentiation in seed dormancy across the range can therefore be considered as a consequence of natural selection in key regulatory genes that manage the hormone signalling pathways responsible for environment-dependent germination (Finch-Savage & Leubner-Metzger, 2006). A decrease in seed dormancy in southern populations can help minimize trade-offs associated with dormancy while enhancing fitness across various life stages (Postma & Ågren, 2022). The strong climate cline in the germination rate is consistent with previous analyses of slow-colonizing forest herb species where an increased germination success in warmer climates was observed (De Frenne et al., 2011; Frenne et al., 2012; Graae et al., 2009). Furthermore, warm-adapted progeny enhanced their juvenile growth rates under drought stress, whereas cold-adapted progeny exhibited reduced juvenile growth rates under drought. This suggests that warm-adapted progeny, which evolved to optimize growth rates in warmer climates (Chaves et al., 2003; Morison & Morecroft, 2008), display increased drought tolerance (Kooyers, 2015). Flowering was significantly delayed in northern populations, which is consistent with the general trend that temperate forest communities tend to flower later in colder climates to prevent frost damage (Flynn & Wolkovich, 2018) and to synchronize their blooming with the emergence of pollinator communities (Memmott et al., 2007). Under drought stress, southern populations further slowed flowering, which is in line with the observation that the flowering season of vernal herbs lag behind under reduced precipitation in order to avoid unsuccessful fruit development under dry conditions (Nam & Kim, 2020).

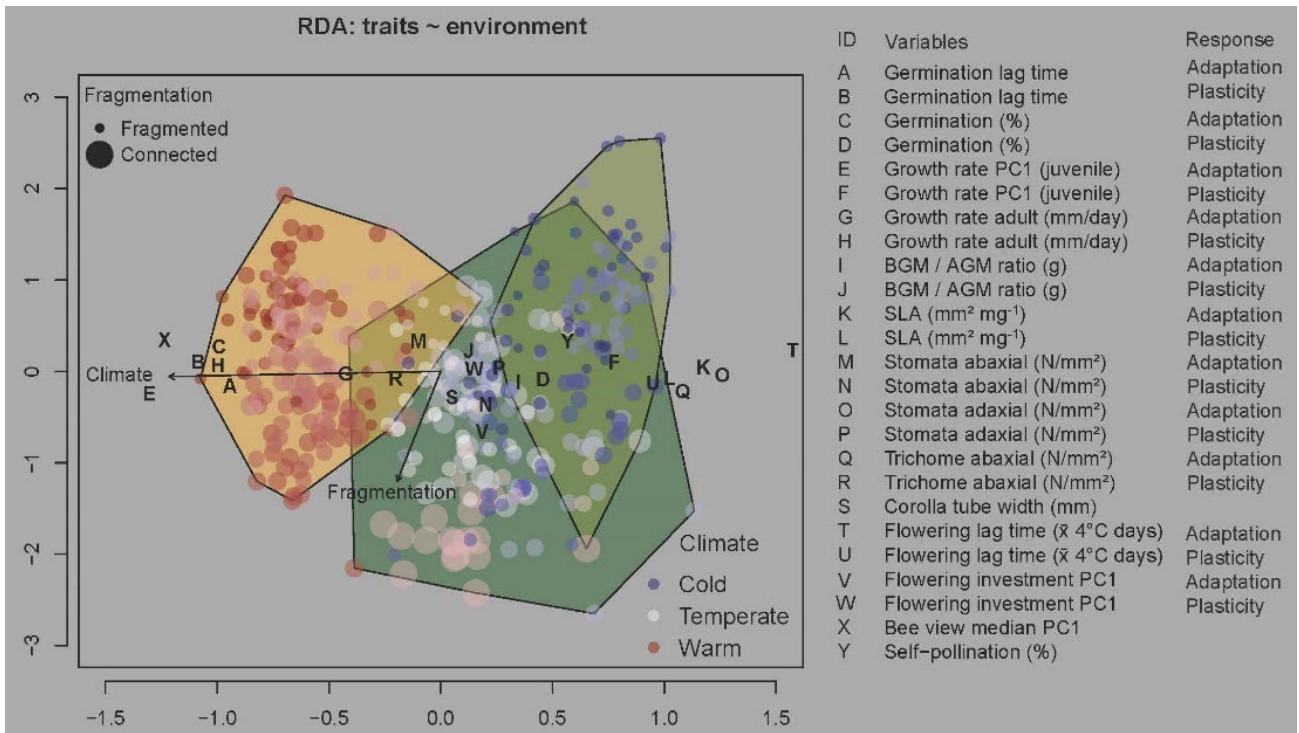

*Figure 6: Redundancy analysis of phenotypic trait variation (Adaptation) and phenotypic plasticity (Plasticity) in* Primula elatior. *Traits were scaled to unit variance as response, and the first PC axis of respectively climate and habitat fragmentation were used as explanatory variables. Each dot indicates a parental origin ID (individuals grouped by F1), colours are defined by the first climate PC axis eigenvalues, the dot size is determined by the eigenvalues of the first habitat fragmentation PC axis. The x axis displays the eigenvalues of the first RDA axis and the y axis displays the eigenvalues of the second RDA axis. The legend for the trait IDs is displayed on the right. The differentiation between phenotypic trait variation (Adaptation) and phenotypic plasticity (Plasticity) is indicated in the response column.*

The overall higher adult growth rate, root:shoot ratio and lower specific leaf area of southern populations is in line with the general observation that herbaceous plants increase relative investment in root biomass and increase leaf density to reduce evaporation and enhance carbon assimilation in warmer climates (Frenne et al., 2012; Lemke et al., 2015; Poorter et al., 2009). Plants, depending on their evolutionary history, will generally display distinct plastic responses to cope with drought stress, aiming to either avoid, escape, or tolerate drought (Kooyers, 2015). In this study, adult growth slowed down, the root:shoot ratio decreased, SLA increased, and abaxial stomata increased under drought stress. Slowing down adult growth indicates a drought avoidance strategy aiming to reduce water loss (Chaves et al., 2003; Whale, 1983). A reduced root:shoot ratio and increased specific leaf area under drought may initially appear counterintuitive, however, by allocating more resources to shoots and expanding leaf area, plants can efficiently maintain photosynthesis in water-limited conditions, a characteristic often observed in shade-tolerant species (McCarthy & Enquist, 2007; Poorter et al., 2012). Increased abaxial stomatal density enables plants to enhance their capacity for gas exchange regulation, which in turn increases their carbon assimilation rate and water use efficiency while maintaining photosynthesis (Xu & Zhou, 2008). The adaptive drought avoidance strategy of southern populations enabled plants to maintain their flowering investment under drought (Karlsson & Méndez, 2005). Conversely, northern populations exhibited an increased root:shoot ratio, along with a reduction in specific leaf area and abaxial stomatal density under drought stress, suggesting an adaptive drought tolerance strategy (Kooyers, 2015; Poorter et al., 2012).

## 4.2. The impact of habitat fragmentation on climate adaptation

In line with our second objective, we found that habitat fragmentation disrupted climate clines and adaptive responses to drought stress in several traits and across life phases. Specifically, southern fragmented populations of *P. elatior* exhibited longer flowering lag times, the adult growth cline was absent in fragmented populations, the relation between climate and the root:shoot ratio was inverted in fragmented populations, and the adaxial trichome density was strongly reduced in southern fragmented populations, leading to strongly reduced flower investment in fragmented populations. Additionally, in fragmented populations, the adaptive capacity to drought stress was impacted across various traits and life stages. Phenotypic plasticity of the growth rate during both juvenile and adult phases, as well as the root:shoot ratio, were reduced in these populations, and plasticity in abaxial stomatal density was completely lacking.

The effective population size plays a crucial role in shaping adaptive differentiation by influencing the interplay between selection and genetic drift. Larger populations are more likely to maintain a higher genetic diversity and experience stronger selection pressures, which in turn fosters adaptive differentiation in response to varying climatic conditions (Leimu & Fischer, 2008). On the other hand, smaller populations are more susceptible to genetic drift, resulting in the random loss of advantageous alleles (Aguilar et al., 2008; Naaf et al., 2021; Pickup et al., 2012). Additionally, habitat fragmentation can lead to reduced gene flow between populations due to an unsuitable landscape matrix and an altered or degraded pollinator community, further reducing genetic diversity and evolutionary potential (Cheptou et al., 2017; Hadley & Betts, 2012; Naaf et al., 2021). The resulting genetic erosion likely led to the observed reduced adaptive potential in several key traits related to development and water regulation and could hinder their ability to cope with climate change. The observed reduction of the glandular trichome density in southern fragmented populations could be an additional consequence of disrupted plant-pollinator networks, potentially leading to a diminished investment in secondary metabolites which attract pollinators and, consequently, decreasing the adaptive advantage of glandular trichomes (Uzelac et al., 2012). As glandular trichomes and the produced secondary metabolites also play a role in protection against UV, drought, and predation this could thus further lead to a reduced stress tolerance (Colombo et al., 2017; Uzelac et al., 2012). Finally, fragmented habitats often experience increased edge effects, leading to altered microclimates (Harper et al., 2005; Hofmeister et al., 2019). Heterogeneous landscapes with a high fragmentation rate also tend to contain more early-successional secondary forests, characterized by distinct tree communities with unique phenology (Dubois & Cheptou, 2017; Vandepitte et al., 2007). Furthermore, small forest fragments often exhibit elevated soil nutrient concentrations due to leaching from surrounding fields, leading to increased aboveground competition (Weathers et al., 2001). These differentiated selection pressures could have led to the slightly faster germination(lag time) and the increased juvenile growth rate enabling populations to occasionally establish in more recent forest stands where interspecific competition is more prevalent (Dubois & Cheptou, 2017; Hermy et al., 1999). Furthermore, phenology of *P. elatior* is highly sensitive to environmental cues and the observed delayed flowering could disrupt plant-pollinator networks in fragmented forests (Baeten et al., 2015). In conclusion, habitat fragmentation disrupts climate clines and adaptive responses to drought stress in *P. elatior*, potentially reducing the long-term resilience and viability of populations. Our study highlights the importance of understanding the complex interplay between habitat fragmentation and species' adaptive potential in the face of ongoing environmental changes. Further untangling the evolutionary factors contributing to diminished climate clines in fragmented habitats will be vital for future research endeavours. Considering habitat fragmentation when evaluating local adaptation not only enhances our understanding of how complex environmental pressures shape adaptive processes but also ensures accurate guidance for conservation efforts aimed at effectively preserving biodiversity.

### 4.3. The role of habitat fragmentation in shaping mating systems

Habitat fragmentation has a notable impact on the mating system of *Primula elatior*, affecting evolutionary selection in multiple flower traits (O3). Flower investment was strongly reduced in fragmented populations along the range, especially in southern populations (Fig. 5). Additionally, southern populations exhibited divergent pollinator signalling (bee view) between fragmented and connected habitats. In these southern fragmented populations, S morphs also demonstrated a marked reduction in herkogamy. Furthermore, the presence of biochemical self-compatibility in fragmented habitats led to a substantial increase in inbreeding coefficients among selfed offspring (Appendix S5). As flowers in fragmented populations likely receive less flower visits from pollinators, investment in flowers becomes less beneficial from an evolutionary viewpoint (Aguilar et al., 2006). Nevertheless, it remains uncertain whether the diminished flower investment stems from disrupted plant-pollinator networks (Jacquemyn et al., 2012), reduced climate clines in fragmented populations (section 4.2), from inbreeding effects (Andersson, 2012), reduced fitness (Lienert, 2004) or other environmental drivers (Weber & Kolb, 2011).

While flower investment was mainly associated with fragmentation, herkogamy, pollinator signalling (bee view), and self-compatibility displayed substantial north to south variation. During the last ice age, the ranges of many forest herbs retracted, with remaining populations residing in temperate regions throughout the Last Glacial Maximum. Postglacial recolonization by dispersal limited forest herbs was particularly slow, still impacting species distributions and genetic imprints to date (Hewitt, 1999; Svenning et al., 2008; Willner et al., 2009). During the northward post-glacial recolonization of Europe many forest herb populations accumulated genetic bottlenecks and underwent strong selection pressures in often isolated founding populations (Koski et al., 2019), leading to a decreased herkogamy and a higher self-compatibility in northern populations (Barrett, 2019; Encinas-Viso et al., 2020). Long-term isolation in glacial refugia and the northward post-glacial recolonization of Europe thus likely exerted strong evolutionary pressures on herkogamy differentiation over extended periods of time (Cain et al., 1998; Cheddadi et al., 2006), but contemporary habitat fragmentation is already leaving its mark on the breakdown of herkogamy in *Primula elatior*. S-morphs, being more susceptible to gravity or wind-induced self-pollination, experience increased selection pressures on herkogamy breakdown, making these effects particularly evident in these morphs. Attraction mechanisms, such as bullseye patterns and UV signalling, can enhance pollinator orientation, and the reduced pollinator availability in fragmented habitats could therefore exert selective pressure on flowers with increased pollinator signalling (Chittka & Wells, 2004; Koski & Ashman, 2014). Pollinator signalling was mainly observed in southern fragmented populations, which may be linked to the biogeographical history along the range or to climate-driven gradients in the pollinator community (Ganuza et al., 2022; Van Daele et al., 2022). Future phylogenetically-controlled analyses could help discern how the increased reflection (higher PC1 values of median UV red, green and blue bands) in fragmented southern populations affects pollinators and how this related to the limited or altered pollinator community in these habitats. In summary, habitat fragmentation considerably influences the evolutionary trajectory of *P. elatior* through its effects on mating system attributes such as reduced flower investment, altered herkogamy, increased inbreeding coefficients among selfed offspring, and evolving flower attraction patterns. Further exploration of these intricate relationships will be vital for informing effective conservation practices.

## 5. Conclusion and implications

*Primula elatior* demonstrated significant climate adaptation across life phases, with southern populations displaying a drought avoidance strategy and northern populations displaying a drought tolerance strategy. These adaptations enable populations to cope with their specific climatic conditions, ensuring their long-term survival and viability. However, habitat fragmentation poses a significant threat to species' climate adaptation and resilience by disrupting adaptive responses to drought stress and altering plant strategies. Fragmented populations showed reduced phenotypic plasticity in key traits related to development and water regulation, potentially hindering their ability to cope with climate change. Furthermore, habitat fragmentation significantly impacted the mating system of *P. elatior*, severely affecting flower investment. The long-term resilience and viability of *P. elatior* populations could be reduced due to eco-evolutionary effects of habitat fragmentation on climate adaptation. Our findings have important implications for conservation management and planning. Understanding the complex interplay between habitat fragmentation and species' adaptive potential is crucial for developing effective strategies to protect and maintain biodiversity in the face of ongoing environmental changes. Conservation efforts should focus on restoring and preserving large, connected habitats, which can enhance gene flow and maintain the genetic diversity necessary for species to adapt to changing environmental conditions. Furthermore, considering habitat fragmentation when evaluating local adaptation in progeny is essential for accurately guiding conservation efforts and ensuring long-term species survival. Future research could further explore the evolutionary drivers of reduced climate clines in fragmented habitats, as well as investigating the effects of habitat fragmentation on other plant species and their capacity to adapt to environmental change. Addressing these knowledge gaps will enable more targeted and effective conservation strategies to be developed, ultimately contributing to the preservation of global biodiversity in a rapidly changing world.

## 6. Acknowledgments

This research was funded by the Flemish Research Foundation (FWO project G091419N). We would like to thank Kasper Van Acker, Jonas Lequeu, Lander Storms, Olivia De Paepe, Niels Duym, Emily Kirsch, Gilles Breugelmans, Jarne Melis, Katlijne Nachtergaele, and Lotte Simons for practical support. Furthermore, we would like to thank TRANSfarm and Meise Botanic Garden for the use of facilities. Finally, we would like to thank Prof. Francis Wyffels for useful comments in regards to the training of the neural networks for the detection of stomata and glandular trichomes.

## 7. Author Contributions

Frederik Van Daele conducted the fieldwork (seed collection and herkogamy data), set up the common garden, measured trait data, developed the image recognition tools, performed the statistical analysis, made the figures, and wrote the initial draft of the manuscript. Olivier Honnay provided feedback on the experimental set-up and on the initial draft. Steven Janssens supervised microscopy of stomata and trichomes and provided feedback on the initial draft. Hanne De Kort conceived the experiment, advised the experimental set-up, and provided feedback on the initial draft.

## 8. Data Availability Statement

Trait data will be deposited in the TRY Plant Trait Database upon publication of the manuscript. A direct link to these datasets will be provided in the final article.

## 9. Conflict of interest

The authors declare no conflict of interest.

## 10. References

Abadi, M., Agarwal, A., Barham, P., Brevdo, E., Chen, Z., Citro, C., Corrado, G. S., Davis, A., Dean, J., Devin, M., Ghemawat, S., Goodfellow, I., Harp, A., Irving, G., Isard, M., Jia, Y., Jozefowicz, R., Kaiser, L., Kudlur, M., … Zheng, X. (2015). *TensorFlow: Large-Scale Machine Learning on Heterogeneous Systems*. https://www.tensorflow.org/

Aguilar, R., Ashworth, L., Galetto, L., & Aizen, M. A. (2006). Plant reproductive susceptibility to habitat fragmentation: Review and synthesis through a meta-analysis. *Ecology Letters*, *9*(8), 968–980. https://doi.org/10.1111/j.1461-0248.2006.00927.x

Aguilar, R., Quesada, M., Ashworth, L., Herrerias-Diego, Y., & Lobo, J. (2008). Genetic

consequences of habitat fragmentation in plant populations: Susceptible signals in plant traits and methodological approaches. *Molecular Ecology*, *17*(24), 5177–5188. https://doi.org/10.1111/j.1365-294X.2008.03971.x

Aiken, L. S., & West, S. G. (1991). *Multiple regression: testing and interpreting interactions*. Sage.

Andersson, S. (2012). Does inbreeding promote evolutionary reduction of flower size? experimental evidence from Crepis tectorum (Asteraceae). *American Journal of Botany*, *99*(8), 1388–1398. https://doi.org/10.3732/ajb.1200116

Armbruster, W. S. (2002). Can indirect selection and genetic context contribute to trait diversification? A transition-probability study of blossom-colour evolution in two genera. *Journal of Evolutionary Biology*, *15*(3), 468–486. https://doi.org/10.1046/j.1420-9101.2002.00399.x

Baeten, L., Sercu, B., Bonte, D., Vanhellemont, M., & Verheyen, K. (2015). Intraspecific variation in flowering phenology affects seed germinability in the forest herb Primula elatior. *Plant Ecology and Evolution*, *148*(2), 283–288. https://doi.org/10.5091/plecevo.2015.1089

Barrett, S. C. H. (2019). 'A most complex marriage arrangement': recent advances on heterostyly and unresolved questions. *New Phytologist*, *224*(3), 1051–1067. https://doi.org/10.1111/nph.16026

Barthelme, S. (2022). *imager: Image Processing Library Based on "CImg."* https://cran.r-project.org/package=imager

Baskin, C. C. (2014). *Seeds: Ecology, Biogeography, And, Evolution of Dormancy and Germination* (2nd ed.). Academic Press.

Baskin, J. M., & Baskin, C. C. (2004). A classification system for seed dormancy. *Seed Science Research*, *14*(July), 1–16. https://doi.org/10.1079/SSR2003150

Bates, D., Mächler, M., Bolker, B., & Walker, S. (2015). Fitting Linear Mixed-Effects Models Using {lme4}. *Journal of Statistical Software*, *67*(1), 1–48. https://doi.org/10.18637/jss.v067.i01

Bawa, K. . S. ., Brys, R., de Crop, E., Hoffmann, M., Jacquemyn, H., Busch, J. W., Schoen, D. J., Hiscock, S. . J., Lloyd, D. G. ., Mansur, L., Gonzalez, M., Rojas, I., Salas, P., Ortiz, M. A., Talavera, S., García-Castaño, J. L., Tremetsberger, K., Stuessy, T., Balao, F., … Tabah, D. a. (2011). Mechanisms of self-incompatibility in flowering plants. *American Journal of Botany*, *66*(1), 1988–2007. https://doi.org/10.1111/j.1558-5646.2011.01505.x

Becklin, K. M., Anderson, J. T., Gerhart, L. M., Wadgymar, S. M., Wessinger, C. A., & Ward, J. K. (2016). Examining plant physiological responses to climate change through an evolutionary lens. *Plant Physiology*, *172*(2), 635–649. https://doi.org/10.1104/pp.16.00793

Bewley, J. D., Bradford, K., & Hilhorst, H. (2012). Dormancy and the Control of Germination. In *Seeds: Physiology of Development, Germination and Dormancy* (pp. 247–279). Springer Science \& Business Media.

Blanquart, F., Kaltz, O., Nuismer, S. L., & Gandon, S. (2013). A practical guide to measuring local adaptation. *Ecology Letters*, *16*(9), 1195–1205. https://doi.org/10.1111/ele.12150

Bonte, D., Van Dyck, H., Bullock, J. M., Coulon, A., Delgado, M., Gibbs, M., Lehouck, V., Matthysen, E., Mustin, K., Saastamoinen, M., Schtickzelle, N., Stevens, V. M., Vandewoestijne, S., Baguette, M., Barton, K., Benton, T. G., Chaput-Bardy, A., Clobert, J., Dytham, C., … Travis, J. M. J. (2012). Costs of dispersal. *Biological Reviews*, *87*(2), 290–312. https://doi.org/10.1111/j.1469-185X.2011.00201.x

Brys, R., & Jacquemyn, H. (2015). Disruption of the distylous syndrome in Primula veris. *Annals of Botany*, *115*(1), 27–39. https://doi.org/10.1093/aob/mcu211

Bucher, S. F., König, P., Menzel, A., Migliavacca, M., Ewald, J., & Römermann, C. (2018). Traits and climate are associated with first flowering day in herbaceous species along elevational gradients. *Ecology and Evolution*, *8*(2), 1147–1158. https://doi.org/10.1002/ece3.3720

Cain, M. L., Damman, H., & Muir, A. (1998). Seed dispersal and the Holocene migration of woodland herbs. *Ecological Monographs*, *68*(3), 325–347. https://doi.org/10.1890/0012-9615(1998)068[0325:SDATHM]2.0.CO;2

Chaves, M. M., Maroco, J. P., & Pereira, J. S. (2003). Understanding plant responses to drought - From genes to the whole plant. *Functional Plant Biology*, *30*(3), 239–264. https://doi.org/10.1071/FP02076

Cheddadi, R., Vendramin, G. G., Litt, T., François, L., Kageyama, M., Lorentz, S., Laurent, J. M., de Beaulieu, J. L., Sadori, L., Jost, A., & Lunt, D. (2006). Imprints of glacial refugia in the modern genetic diversity of Pinus sylvestris. *Global Ecology and Biogeography*, *15*(3), 271–282. https://doi.org/10.1111/j.1466-822X.2006.00226.x

Chen, H. (2018). *VennDiagram: Generate High-Resolution Venn and Euler Plots*. https://cran.r-project.org/package=VennDiagram

Cheptou, P. O., Hargreaves, A. L., Bonte, D., & Jacquemyn, H. (2017). Adaptation to fragmentation: Evolutionary dynamics driven by human influences. *Philosophical Transactions of the Royal Society B: Biological Sciences*, *372*(1712). https://doi.org/10.1098/rstb.2016.0037

Chittka, L., & Wells, H. (2004). Color vision in bees: mechanisms, ecology, and evolution. *How Simple Nervous Systems Create Complex Perceptual Worlds*, 165–191.

Chollet, F., & others. (2015). *Keras*. GitHub. https://github.com/fchollet/keras

Chubaty, A. M., & McIntire, E. J. B. (2022). *SpaDES: Develop and Run Spatially Explicit Discrete Event Simulation Models*. https://cran.r-project.org/package=SpaDES

Cleland, E. E., Isabelle, C., & Annette, M. (2007). Shifing Plant Phenology In Response to Global Change. *Trends in Ecology & Evolution*, *22*(7), 357–365.

Colombo, P. S., Flamini, G., Rodondi, G., Giuliani, C., Santagostini, L., & Fico, G. (2017). Phytochemistry of European Primula species. *Phytochemistry*, *143*, 132–144. https://doi.org/10.1016/j.phytochem.2017.07.005

Coors, A., & DeMeester, L. (2009). Synergistic, antagonistic and additive effects of multiple stressors: Predation threat, parasitism and pesticide exposure in Daphnia magna. *Journal of Applied Ecology*, *46*(5), 1138. https://doi.org/10.1111/j.1365-2664.2009.01727-1.x

Copernicus Land Monitoring Service. (2018). *Forest Type* (HRL_ForestType_2018). European Environment Agency. https://land.copernicus.eu/pan-european/high-resolution-layers/forests/forest-type-1/status-maps/forest-type-2018?tab=mapview

Cote, J., Bestion, E., Jacob, S., Travis, J., Legrand, D., & Baguette, M. (2017). Evolution of dispersal strategies and dispersal syndromes in fragmented landscapes. *Ecography*, *40*(1), 56–73. https://doi.org/10.1111/ecog.02538

Dawson, J. F., & Richter, A. W. (2006). Probing three-way interactions in moderated multiple regression: Development and application of a slope difference test. *Journal of Applied Psychology*, *91*(4), 917–926. https://doi.org/10.1037/0021-9010.91.4.917

De Frenne, P., Brunet, J., Shevtsova, A., Kolb, A., Graae, B. J., Chabrerie, O., Cousins, S. A., Decocq, G., De Schrijver, A., Diekmann, M., Gruwez, R., Heinken, T., Hermy, M., Nilsson, C., Stanton, S., Tack, W., Willaert, J., & Verheyen, K. (2011). Temperature effects on forest herbs assessed by warming and transplant experiments along a latitudinal

gradient. *Global Change Biology*, *17*(10), 3240–3253. https://doi.org/10.1111/j.1365-2486.2011.02449.x

De Frenne, P., Graae, B. J., Rodríguez-Sánchez, F., Kolb, A., Chabrerie, O., Decocq, G., De Kort, H., De Schrijver, A., Diekmann, M., Eriksson, O., Gruwez, R., Hermy, M., Lenoir, J., Plue, J., Coomes, D. A., & Verheyen, K. (2013). Latitudinal gradients as natural laboratories to infer species' responses to temperature. *Journal of Ecology*, *101*(3), 784–795. https://doi.org/10.1111/1365-2745.12074

De Kort, H., Panis, B., Helsen, K., Douzet, R., Janssens, S. B., & Honnay, O. (2020). Pre-adaptation to climate change through topography-driven phenotypic plasticity. *Journal of Ecology*, *108*(4), 1465–1474. https://doi.org/10.1111/1365-2745.13365

De Witte, L. C., & Stöcklin, J. (2010). Longevity of clonal plants: Why it matters and how to measure it. *Annals of Botany*, *106*(6), 859–870. https://doi.org/10.1093/aob/mcq191

Dickinson, H. G. (1990). *Self-incompatibility in flowering plants: Evolution, Diversity, and Mechanisms* (V. E. Franklin-Tong (ed.); Vol. 12, Issue 4). Springer. https://doi.org/10.1002/bies.950120403

Dubois, J., & Cheptou, P. O. (2017). Effects of fragmentation on plant adaptation to urban environments. *Philosophical Transactions of the Royal Society B: Biological Sciences*, *372*(1712). https://doi.org/10.1098/rstb.2016.0038

Dunnington, D., & Harvey, P. (2021). *exifr: EXIF Image Data in R*. https://cran.r-project.org/package=exifr

Encinas-Viso, F., Young, A. G., & Pannell, J. R. (2020). The loss of self-incompatibility in a range expansion. *Journal of Evolutionary Biology*, *33*(9), 1235–1244. https://doi.org/10.1111/jeb.13665

Endels, P., Adriaens, D., Verheyen, K., & Hermy, M. (2004). Population structure and adult plant performance of forest herbs in three contrasting habitats. *Ecography*, *27*(2), 225–241. https://doi.org/10.1111/j.0906-7590.2004.03731.x

Eriksson, O., & Ehrlen, J. (2001). *Landscape fragmentation and the viability of plant populations* (J. Silvertown (ed.)). Blackwell.

Etterson, J. R., & Shaw, R. G. (2001). Constraint to adaptive evolution in response to global warming. *Science*, *294*(5540), 151–154. https://doi.org/10.1126/science.1063656

Fick, S. E., & Hijmans, R. J. (2017). WorldClim 2: new 1-km spatial resolution climate surfaces for global land areas. *International Journal of Climatology*, *37*(12), 4302–4315. https://doi.org/10.1002/joc.5086

Finch-Savage, W. E., & Leubner-Metzger, G. (2006). Seed dormancy and the control of germination. In *New Phytologist* (Vol. 171, Issue 3, pp. 501–523). https://doi.org/10.1111/j.1469-8137.2006.01787.x

Flanders Research Foundation. (2022). *Flemish Supercomputer Center*. https://www.vscentrum.be

Flynn, D. F. B., & Wolkovich, E. M. (2018). Temperature and photoperiod drive spring phenology across all species in a temperate forest community. *New Phytologist*, *219*(4), 1353–1362. https://doi.org/10.1111/nph.15232

Franks, S. J., Weber, J. J., & Aitken, S. N. (2014). Evolutionary and plastic responses to climate change in terrestrial plant populations. *Evolutionary Applications*, *7*(1), 123–139. https://doi.org/10.1111/eva.12112

Frenne, P. De, Graae, B. J., Brunet, J., Shevtsova, A., Schrijver, A. De, Chabrerie, O., Cousins, S. A. O., Decocq, G., Diekmann, M., Hermy, M., Heinken, T., Kolb, A., Nilsson, C., Stanton, S., & Verheyen, K. (2012). The response of forest plant regeneration to temperature variation along a latitudinal gradient. *Annals of Botany*, *109*(5), 1037–1046. https://doi.org/10.1093/aob/mcs015

Ganuza, C., Redlich, S., Uhler, J., Tobisch, C.,

Rojas-Botero, S., Peters, M. K., Zhang, J., Benjamin, C. S., Englmeier, J., Ewald, J., Fricke, U., Haensel, M., Kollmann, J., Riebl, R., Uphus, L., Müller, J., & Steffan-Dewenter, I. (2022). Interactive effects of climate and land use on pollinator diversity differ among taxa and scales. *Science Advances*, *8*(18). https://doi.org/10.1126/sciadv.abm9359

Gómez-Martínez, C., Aase, A. L. T. O., Totland, Ø., Rodríguez-Pérez, J., Birkemoe, T., Sverdrup-Thygeson, A., & Lázaro, A. (2020). Forest fragmentation modifies the composition of bumblebee communities and modulates their trophic and competitive interactions for pollination. *Scientific Reports*, *10*(1), 1–15. https://doi.org/10.1038/s41598-020-67447-y

Gower, J. C. (1971). A General Coefficient of Similarity and Some of Its Properties. *Biometrics*, *27*(4), 857. https://doi.org/10.2307/2528823

Graae, B. J., Verheyen, K., Kolb, A., Van Der Veken, S., Heinken, T., Chabrerie, O., Diekmann, M., Valtinat, K., Zindel, R., Karlsson, E., Ström, L., Decocq, G., Hermy, M., & Baskin, C. C. (2009). Germination requirements and seed mass of slow- And fast colonizing temperate forest herbs along a latitudinal gradient. *Ecoscience*, *16*(2), 248–257. https://doi.org/10.2980/16-2-3234

Gumbert, A. (2000). Color choices by bumble bees (Bombus terrestris): innate preferences and generalization after learning. *Behavioral Ecology and Sociobiology*, *48*(1), 36–43. http://link.springer.com/10.1007/s002650000213

Hadley, A. S., & Betts, M. G. (2012). The effects of landscape fragmentation on pollination dynamics: Absence of evidence not evidence of absence. *Biological Reviews*, *87*(3), 526–544. https://doi.org/10.1111/j.1469-185X.2011.00205.x

Hanski, I., Mononen, T., & Ovaskainen, O. (2011). Eco-evolutionary metapopulation dynamics and the spatial scale of adaptation. *American Naturalist*, *177*(1), 29–43. https://doi.org/10.1086/657625

Harper, K. A., Macdonald, S. E., Burton, P. J., Chen, J., Brosofske, K. D., Saunders, S. C., Euskirchen, E. S., Roberts, D., Jaiteh, M. S., & Esseen, P. A. (2005). Edge influence on forest structure and composition in fragmented landscapes. In *Conservation Biology* (Vol. 19, Issue 3, pp. 768–782). https://doi.org/10.1111/j.1523-1739.2005.00045.x

Hartig, F. (2022). *DHARMa: Residual Diagnostics for Hierarchical (Multi-Level / Mixed) Regression Models*. https://cran.r-project.org/package=DHARMa

He, H. S., DeZonia, B. E., & Mladenoff, D. J. (2001). An aggregation index (AI) to quantify spatial patterns of landscapes. *Landscape Ecology*, *16*(1), 87. https://doi.org/10.1023/A:1017308405507

Herlihy, C. R., & Eckert, C. G. (2007). Evolutionary analysis of a key floral trait in Aquilegia canadensis (Ranunculaceae): Genetic variation in herkogamy and its effect on the mating system. *Evolution*, *61*(7), 1661–1674. https://doi.org/10.1111/j.1558-5646.2007.00137.x

Hermy, M., Honnay, O., Firbank, L., Grashof-Bokdam, C., & Lawesson, J. E. (1999). An ecological comparison between ancient and other forest plant species of Europe, and the implications for forest conservation. *Biological Conservation*, *91*(1), 9–22. https://doi.org/10.1016/S0006-3207(99)00045-2

Hesselbarth, M. H. K., Sciaini, M., With, K. A., Wiegand, K., & Nowosad, J. (2019). landscapemetrics : an open-source R tool to calculate landscape metrics . *Ecography*, 1–10. https://doi.org/10.1111/ecog.04617

Hewitt, G. M. (1999). Post-glacial re-colonization of European biota. *Biological Journal of the Linnean Society*, *68*(1–2), 87–112. https://doi.org/10.1006/bijl.1999.0332

Hijmans, R. J. (2019). *raster: Geographic Data Analysis and Modeling. R package version 3.0-7.* https://cran.r-project.org/package=raster

Hofmeister, J., Hošek, J., Brabec, M., Stŕalková, R., Mýlová, P., Bouda, M., Pettit, J. L., Rydval, M., & Svoboda, M. (2019). Microclimate edge effect in small fragments of temperate forests in the context of climate change. *Forest Ecology and Management*, *448*(May), 48–56. https://doi.org/10.1016/j.foreco.2019.05.069

Honnay, O., Butaye, J., Jacquemyn, H., Verheyen, K., Hermy, M., Bossuyt, B., Butaye, J., Jacquemyn, H., Bossuyt, B., Hermy, M., Verheyen, K., Hermy, M., & Bossuyt, B. (2002). Possible effects of habitat fragmentation and climate change on the range of forest plant species. *Ecology Letters*, *5*(4), 525–530. https://doi.org/10.1046/j.1461-0248.2002.00346.x

Jacquemyn, H., Brys, R., & Hermy, M. (2002). Patch occupancy, population size and reproductive success of a forest herb (Primula elatior) in a fragmented landscape. *Oecologia*, *130*(4), 617–625. https://doi.org/10.1007/s00442-001-0833-0

Jacquemyn, H., de Meester, L., Jongejans, E., & Honnay, O. (2012). Evolutionary changes in plant reproductive traits following habitat fragmentation and their consequences for population fitness. *Journal of Ecology*, *100*(1), 76–87. https://doi.org/10.1111/j.1365-2745.2011.01919.x

Jaeger, J. A. G. (2000). Landscape division, splitting index, and effective mesh size: New measures of landscape fragmentation. *Landscape Ecology*, *15*(2), 115–130. https://doi.org/10.1023/A:1008129329289

Jump, A. S., & Peñuelas, J. (2005). Running to stand still: Adaptation and the response of plants to rapid climate change. *Ecology Letters*, *8*(9), 1010–1020. https://doi.org/10.1111/j.1461-0248.2005.00796.x

Karlsson, P. S., & Méndez, M. (2005). The Resource Economy of Plant Reproduction. In E. Reekie & F. A. Bazzaz (Eds.), *Reproductive Allocation in Plants*. Elsevier. https://www.elsevier.com/books/reproductive-allocation-in-plants/reekie/978-0-12-088386-8

Katabuchi, & Masatoshi. (2015). LeafArea: An R Package for Rapid Digital Image Analysis of Leaf Area. *Ecological Research*, *30*(6), 1073–1077. https://doi.org/10.1007/s11284-015-1307-x

Kawecki, T. J., & Ebert, D. (2004). Conceptual issues in local adaptation. *Ecology Letters*, *7*(12), 1225–1241. https://doi.org/10.1111/j.1461-0248.2004.00684.x

Keitt, T. H., Urban, D. L., & Milne, B. T. (1997). Detecting Critical Scales in Fragmented Landscapes. *Conservation Ecology*, *1*(1), 1–13. https://eur-lex.europa.eu/legal-content/PT/TXT/PDF/?uri=CELEX:32016R0679&from=PT%0Ahttp://eur-lex.europa.eu/LexUriServ/LexUriServ.do?uri=CELEX:52012PC0011:pt:NOT

Kooyers, N. J. (2015). The evolution of drought escape and avoidance in natural herbaceous populations. *Plant Science*, *234*, 155–162. https://doi.org/10.1016/j.plantsci.2015.02.012

Koski, M. H., & Ashman, T. L. (2013). Quantitative variation, heritability, and trait correlations for ultraviolet floral traits in Argentina anserina (Rosaceae): Implications for floral evolution. *Evolution*, *174*(8), 1109–1120. https://doi.org/10.1086/671803

Koski, M. H., & Ashman, T. L. (2014). Dissecting pollinator responses to a ubiquitous ultraviolet floral pattern in the wild. *Functional Ecology*, *28*(4), 868–877. https://doi.org/10.1111/1365-2435.12242

Koski, M. H., Layman, N. C., Prior, C. J., Busch, J. W., & Galloway, L. F. (2019). Selfing ability and drift load evolve with range expansion. *Evolution Letters*, *3*(5), 500–512. https://doi.org/10.1002/evl3.136

Kwak, M. M., Velterop, O., & van Andel, J. (1998). Pollen and gene flow in fragmented habitats. *Applied Vegetation Science*, *1*(1), 37–54. https://doi.org/10.2307/1479084

Lai, J., Zou, Y., Zhang, J., & Peres-Neto, P. R.

(2022). Generalizing hierarchical and variation partitioning in multiple regression and canonical analyses using the rdacca.hp R package. *Methods in Ecology and Evolution*, *13*(4), 782–788. https://doi.org/10.1111/2041-210X.13800

Leimu, R., & Fischer, M. (2008). A meta-analysis of local adaptation in plants. *PLoS ONE*, *3*(12), 1–8. https://doi.org/10.1371/journal.pone.0004010

Leimu, R., Vergeer, P., Angeloni, F., & Ouborg, N. J. (2010). Habitat fragmentation, climate change, and inbreeding in plants. *Annals of the New York Academy of Sciences*, *1195*(May), 84–98. https://doi.org/10.1111/j.1749-6632.2010.05450.x

Lemke, I. H., Kolb, A., Graae, B. J., De Frenne, P., Acharya, K. P., Blandino, C., Brunet, J., Chabrerie, O., Cousins, S. A. O., Decocq, G., Heinken, T., Hermy, M., Liira, J., Schmucki, R., Shevtsova, A., Verheyen, K., & Diekmann, M. (2015). Patterns of phenotypic trait variation in two temperate forest herbs along a broad climatic gradient. *Plant Ecology*, *216*(11), 1523–1536. https://doi.org/10.1007/s11258-015-0534-0

Lienert, J. (2004). Habitat fragmentation effects of fitness of plant populations - A review. *Journal for Nature Conservation*, *12*(1), 53–72. https://doi.org/10.1016/j.jnc.2003.07.002

Liu, C., He, N., Zhang, J., Li, Y., Wang, Q., Sack, L., & Yu, G. (2018). Variation of stomatal traits from cold temperate to tropical forests and association with water use efficiency. *Functional Ecology*, *32*(1), 20–28. https://doi.org/10.1111/1365-2435.12973

Liu, W., Zheng, L., & Qi, D. (2020). Variation in leaf traits at different altitudes reflects the adaptive strategy of plants to environmental changes. *Ecology and Evolution*, *10*(15), 8166–8175. https://doi.org/10.1002/ece3.6519

Lortie, C. J., & Hierro, J. L. (2022). A synthesis of local adaptation to climate through reciprocal common gardens. *Journal of Ecology*, *110*(5), 1015–1021. https://doi.org/10.1111/1365-2745.13664

Lüdecke, D. (2021). *sjPlot: Data Visualization for Statistics in Social Science*. https://cran.r-project.org/package=sjPlot

Lüdecke, D., Ben-Shachar, M. S., Patil, I., Waggoner, P., & Makowski, D. (2021). performance: An R Package for Assessment, Comparison and Testing of Statistical Models. *Journal of Open Source Software*, *6*(60), 3139. https://doi.org/10.21105/joss.03139

Lunau, K., Verhoeven, C., & Ren, Z.-X. (2018). False-colour photography: a novel digital approach to visualize the bee view of flowers. *Journal of Pollination Ecology*, *23*(12), 102–118. https://doi.org/https://doi.org/10.26786/1920-7603(2018)11

Maechler, M., Rousseeuw, P., Struyf, A., Hubert, M., & Hornik, K. (2017). *cluster: Cluster Analysis Basics and Extensions* (R package version 2.0.6.).

McCarthy, M. C., & Enquist, B. J. (2007). Consistency between an allometric approach and optimal partitioning theory in global patterns of plant biomass allocation. *Functional Ecology*, *21*(4), 713–720. https://doi.org/10.1111/j.1365-2435.2007.01276.x

Meeus, S., Van den Bulcke, J., & wyffels, F. (2020). From leaf to label: A robust automated workflow for stomata detection. *Ecology and Evolution*, *10*(17), 9178–9191. https://doi.org/10.1002/ece3.6571

Memmott, J., Craze, P. G., Waser, N. M., & Price, M. V. (2007). Global warming and the disruption of plant-pollinator interactions. *Ecology Letters*, *10*(8), 710–717. https://doi.org/10.1111/j.1461-0248.2007.01061.x

Mitchell, R. J., Ashman, T., The, S., Phytologist, N., Leake, J. R., Cameron, D. D., & Beerling, D. J. (2008). Predicting Evolutionary Consequences of Pollinator Declines : The Long and Short of Floral Evolution. *New*

*Phytologist*, *177*, 576–579.

Morison, J. I. L., & Morecroft, M. D. (2008). *Plant growth and climate change*. John Wiley & Sons.

Murrell, P. (2019). *hexView: Viewing Binary Files*. https://cran.r-project.org/package=hexView

Murtagh, F., & Legendre, P. (2014). Ward's Hierarchical Agglomerative Clustering Method: Which Algorithms Implement Ward's Criterion? *Journal of Classification*, *31*(3), 274–295. https://doi.org/10.1007/s00357-014-9161-z

Naaf, T., Feigs, J. T., Huang, S., Brunet, J., Cousins, S. A. O., Decocq, G., De Frenne, P., Diekmann, M., Govaert, S., Hedwall, P. O., Helsen, K., Lenoir, J., Liira, J., Meeussen, C., Plue, J., Poli, P., Spicher, F., Vangansbeke, P., Vanneste, T., … Kramp, K. (2021). Sensitivity to habitat fragmentation across European landscapes in three temperate forest herbs. *Landscape Ecology*, *36*(10), 2831–2848. https://doi.org/10.1007/s10980-021-01292-w

Naaf, T., Feigs, J. T., Huang, S., Brunet, J., Cousins, S. A. O., Decocq, G., De Frenne, P., Diekmann, M., Govaert, S., Hedwall, P. O., Lenoir, J., Liira, J., Meeussen, C., Plue, J., Vangansbeke, P., Vanneste, T., Verheyen, K., Holzhauer, S. I. J., & Kramp, K. (2021). Context matters: the landscape matrix determines the population genetic structure of temperate forest herbs across Europe. *Landscape Ecology*, *7*. https://doi.org/10.1007/s10980-021-01376-7

Nam, B. E., & Kim, J. G. (2020). Flowering season of vernal herbs is shortened at elevated temperatures with reduced precipitation in early spring. *Scientific Reports*, *10*(1), 1–10. https://doi.org/10.1038/s41598-020-74566-z

Oksanen, J., Blanchet, F. G., Kindt, R., Legendre, P., Minchin, P. R., O'Hara, R. B., Simpson, G. L., Solymos, P., Stevens, M. H. H., Wagner, H., Friendly, M., Kindt, R., Legendre, P., McGlinn, D., Minchin, P. R., O'Hara, R. B., Simpson, G. L., Solymos, P., Stevens, M. H. H., … Wagner, H. (2013). *vegan: Community Ecology Package* (V. 2.5-7). https://cran.r-project.org/package=vegan

Olivieri, I., Michalakis, Y., & Gouyon, P. H. (1995). Metapopulation genetics and the evolution of dispersal. *American Naturalist*, *146*(2), 202–228. https://doi.org/10.1086/285795

Ooms, J. (2021). *magick: Advanced Graphics and Image-Processing in R*. https://cran.r-project.org/package=magick

Opedal, Ø. H. (2019). The evolvability of animal-pollinated flowers: towards predicting adaptation to novel pollinator communities. *New Phytologist*, *221*(2), 1128–1135. https://doi.org/10.1111/nph.15403

Opedal, Ø. H., Bolstad, G. H., Hansen, T. F., Armbruster, W. S., & Pélabon, C. (2017). The evolvability of herkogamy: Quantifying the evolutionary potential of a composite trait. *Evolution*, *71*(6), 1572–1586. https://doi.org/10.1111/evo.13258

Pickup, M., Field, D. L., Rowell, D. M., & Young, A. G. (2012). Predicting local adaptation in fragmented plant populations: Implications for restoration genetics. *Evolutionary Applications*, *5*(8), 913–924. https://doi.org/10.1111/j.1752-4571.2012.00284.x

Poorter, H., Niinemets, Ü., Poorter, L., Wright, I. J., & Villar, R. (2009). Causes and consequences of variation in leaf mass per area (LMA): A meta-analysis. In *New Phytologist* (Vol. 182, Issue 3, pp. 565–588). https://doi.org/10.1111/j.1469-8137.2009.02830.x

Poorter, H., Niklas, K. J., Reich, P. B., Oleksyn, J., Poot, P., & Mommer, L. (2012). Biomass allocation to leaves, stems and roots: Meta-analyses of interspecific variation and environmental control. In *New Phytologist* (Vol. 193, Issue 1, pp. 30–50). https://doi.org/10.1111/j.1469-8137.2011.03952.x

Porcher, E., & Lande, R. (2005). The evolution of self-fertilization and inbreeding depression under pollen discounting and pollen

limitation. *Journal of Evolutionary Biology*, *18*(3), 497–508. https://doi.org/10.1111/j.1420-9101.2005.00905.x

Postma, F. M., & Ågren, J. (2022). Effects of primary seed dormancy on lifetime fitness of Arabidopsis thaliana in the field. *Annals of Botany*, *129*(7), 795–808. https://doi.org/10.1093/aob/mcac010

Qi, Y., Wei, W., Chen, C., & Chen, L. (2019). Plant root-shoot biomass allocation over diverse biomes: A global synthesis. *Global Ecology and Conservation*, *18*(18), e00606. https://doi.org/10.1016/j.gecco.2019.e00606

Raiche, G., & Magis, D. (2020). *nFactors: Parallel Analysis and Other Non Graphical Solutions to the Cattell Scree Test*. https://cran.r-project.org/package=nFactors

Revelle, W. (2022). *psych: Procedures for Psychological, Psychometric, and Personality Research*. https://cran.r-project.org/package=psych

Richardson, A. D., Bailey, A. S., Denny, E. G., Martin, C. W., & O'Keefe, J. (2006). Phenology of a northern hardwood forest canopy. *Global Change Biology*, *12*(7), 1174–1188. https://doi.org/10.1111/j.1365-2486.2006.01164.x

Rosbakh, S., Römermann, C., & Poschlod, P. (2015). Specific leaf area correlates with temperature: new evidence of trait variation at the population, species and community levels. *Alpine Botany*, *125*(2), 79–86. https://doi.org/10.1007/s00035-015-0150-6

Schou, O. (1983). The distyly in Primula elatior (L.) Hill (Primulaceae), with a study of flowering phenology and pollen flow. *Botanical Journal of the Linnean Society*, *86*(3), 261–274. https://doi.org/10.1111/j.1095-8339.1983.tb00973.x

Schumaker, N. H. (1996). Using landscape indices to predict habitat connectivity. *Ecology*, *77*(4), 1210–1225. https://doi.org/10.2307/2265590

Skov, F., & Svenning, J. C. (2004). Potential impact of climatic change on the distribution of forest herbs in Europe. *Ecography*, *27*(3), 366–380. https://doi.org/10.1111/j.0906-7590.2004.03823.x

Strauss, S. Y., & Whittall, J. B. (2006). Non-pollinator agents of selection on floral traits. In *Ecology and evolution of flowers* (Vol. 208, Issue January 2006, pp. 120–138).

Svenning, J.-C., Normand, S., & Skov, F. (2008). Postglacial Dispersal Limitation of Widespread Forest Plant Species in Nemoral Europe. *Ecography*, *31*(3), 316–326. https://doi.org/10.1111/j.2008.0906-7590.05206.x

Taylor, K., Woodell, S. R. J. J., Taylort, K., & Woodellt, S. R. J. (2008). Biological Flora of the British Isles: Primula elatior (L.) Hill. *Journal of Ecology*, *96*(5), 1098–1116. https://doi.org/10.1111/j.1365-2745.2008.01418.x

Uzelac, B., Stojičić, D., Budimir, S., Ramawat, K. G., Ekiert, H. M., & Goyal, S. (2012). Glandular Trichomes on the Leaves of Nicotiana tabacum: Morphology, Developmental Ultrastructure, and Secondary Metabolites. In K. G. Ramawat, H. M. Ekiert, & S. Goyal (Eds.), *Plant Cell and Tissue Differentiation and Secondary Metabolites: Vol. 23(2)*. Springer.

Van Daele, F., Honnay, O., & De Kort, H. (2021). The role of dispersal limitation and reforestation in shaping the distributional shift of a forest herb under climate change. *Diversity and Distributions*, *27*(9), 1775–1791. https://doi.org/10.1111/ddi.13367

Van Daele, F., Honnay, O., & De Kort, H. (2022). Genomic analyses point to a low evolutionary potential of prospective source populations for assisted migration in a forest herb. *Evolutionary Applications*, *2022*(00), 1–16. https://doi.org/10.1111/eva.13485

Van Rossum, F. (2008). Conservation of long-lived perennial forest herbs in an urban context: Primula elatior as study case. *Conservation Genetics*, *9*(1), 119–128.

https://doi.org/10.1007/s10592-007-9314-2

Vandepitte, K., Jacquemyn, H., Roldán-Ruiz, I., & Honnay, O. (2007). Landscape genetics of the self-compatible forest herb Geum urbanum: effects of habitat age, fragmentation and local environment. *Molecular Ecology*, *16*, 4171–4179. https://doi.org/10.1111/j.1365-294X.2007.03473.x

Verheyen, K., Honnay, O., Motzkin, G., Hermy, M., & Foster, D. R. (2003). Response of forest plant species to land-use change: A life-history trait-based approach. *Journal of Ecology*, *91*(4), 563–577. https://doi.org/10.1046/j.1365-2745.2003.00789.x

Vittoz, P., & Engler, R. (2007). Seed dispersal distances: A typology based on dispersal modes and plant traits. *Botanica Helvetica*, *117*(2), 109–124. https://doi.org/10.1007/s00035-007-0797-8

Weathers, K. C., Cadenasso, M. L., & Pickett, S. T. A. (2001). Forest edges as nutrient and pollutant concentrators: Potential synergisms between fragmentation, forest canopies, and the atmosphere. *Conservation Biology*, *15*(6), 1506–1514. https://doi.org/10.1046/j.1523-1739.2001.01090.x

Weber, A., & Kolb, A. (2011). Evolutionary consequences of habitat fragmentation: Population size and density affect selection on inflorescence size in a perennial herb. *Evolutionary Ecology*, *25*(2), 417–428. https://doi.org/10.1007/s10682-010-9430-1

Whale, D. M. (1983). The Response of Primula Species to Soil Waterlogging and Soil Drought. *International Association for Ecology*, *58*(2), 272–277.

Wickham, H. (2016). *ggplot2: Elegant Graphics for Data Analysis*. Springer-Verlag New York. https://ggplot2.tidyverse.org

Wickham, H. (2021). *tidyr: Tidy Messy Data*. https://cran.r-project.org/package=tidyr

Willner, W., Di Pietro, R., & Bergmeier, E. (2009). Phytogeographical evidence for post-glacial dispersal limitation of European beech forest species. *Ecography*, *32*(6), 1011–1018. https://doi.org/10.1111/j.1600-0587.2009.05957.x

Xu, Z., & Zhou, G. (2008). Responses of leaf stomatal density to water status and its relationship with photosynthesis in a grass. *Journal of Experimental Botany*, *59*(12), 3317–3325. https://doi.org/10.1093/jxb/ern185

Zhang, D. (2021). *rsq: R-Squared and Related Measures* (V. 2.2). https://cran.r-project.org/package=rsq

Zhang, W., Hu, Y. F., He, X., Zhou, W., & Shao, J. W. (2021). Evolution of Autonomous Selfing in Marginal Habitats: Spatiotemporal Variation in the Floral Traits of the Distylous Primula wannanensis. *Frontiers in Plant Science*, *12*(December), 1–15. https://doi.org/10.3389/fpls.2021.781281